\documentstyle[preprint,revtex]{aps}
\begin{document}
\draft
\begin{title}
Predictions for Impurity-Induced $T_c$ Suppression\\
in the High-Temperature Superconductors
\end{title}

\author{R. J. Radtke and K. Levin}
\begin{instit}
Department of Physics and the James Franck Institute,\\
The University of Chicago,
Chicago, Illinois  60637
\end{instit}
\author{H.-B. Sch\"uttler}
\begin{instit}
Department of Physics and Astronomy,
The University of Georgia,
Athens, Georgia  30602
\end{instit}
\author{M. R. Norman}
\begin{instit}
Materials Science Division,
Argonne National Laboratory,
Argonne, Illinois  60439
\end{instit}

\begin{abstract}

We address the question of whether
anisotropic superconductivity is compatible
with the evidently weak sensitivity
of the critical temperature $T_c$ to sample quality in the high-$T_c$
copper oxides.
We examine this issue quantitatively by solving
the strong-coupling Eliashberg
equations numerically as well as analytically for
s-wave impurity scattering within the second Born approximation.
For pairing interactions
with a characteristically low energy scale, we find an
approximately universal dependence of the d-wave superconducting
transition temperature on the planar residual resistivity
which is independent of the details of the microscopic pairing.
These results, in conjunction with
future systematic experiments, should help elucidate the
symmetry of the order parameter in the cuprates.

\end{abstract}

\narrowtext

A growing body of experimental evidence has
been interpreted as supporting an anisotropic pairing
state in the high-temperature superconductors \cite{dExp}.
There are indications, however,
that the measured values of the superconducting critical temperature
$T_c$ do not depend strongly on sample quality.
Because both magnetic and non-magnetic impurities
are pair-breaking in anisotropic
superconductors \cite{MSV}, this latter statement
appears to be incompatible with anisotropic pairing.

Two arguments can be offered to explain this
apparent contradiction, but neither is quantitative enough to settle
the issue.
First, the coherence lengths $\xi \approx$ 10 \AA $\;$ in the cuprates,
while the mean free paths $l \approx$ 100 - 200 \AA $\;$\cite{Batlogg}.
Since one expects $T_c$ to be affected by impurities
only when $\xi \sim l$, the small ratio
$\xi / l \approx $ 0.1 suggests that $T_c$ should not be sensitive
to defects.
However, for magnetic impurities
in s-wave superconductors, superconductivity is completely
destroyed for $\xi / l \cong$ 0.12 - 0.17 \cite{AG}, implying that
$\xi / l$ in the cuprates may not be small enough to avoid
an impurity-induced reduction of $T_c$.
Second, one observes significant inelastic scattering in
these materials which could act to mask impurity effects.
Inelastic scattering would only
be able to screen the defects effectively if the
inelastic mean free path were
much shorter than the impurity mean free path.
Estimates of these quantities \cite{Batlogg,Orenstein}
indicate that they are comparable in high-quality samples.

In this paper, we examine quantitatively the effect of impurities
in the cuprates by generalizing the
Abrikosov-Gor'kov scaling law \cite{AG} to anisotropic
strong-coupling superconductors.
We present model-independent
predictions of $T_c$ as a function of planar residual resistivity
due to non-magnetic impurities for a 90 K d-wave
superconductor and compare these results to the
response of an s-wave superconductor.
We also check the validity of this analytical result by calculating
numerically the suppression of $T_c$
induced by structureless impurities treated within the second
Born approximation and in strong-coupling Eliashberg theory.
Finally, we discuss how these predictions may be tested experimentally.
Other authors have examined this question using
analytical \cite{MSV} as well as numerical techniques
\cite{MBP,Lenck,Hotta}.
The goal of the present
work, however, is to quantify the expected $T_c$
suppression for d-wave models of the high temperature superconductors,
so that the d-wave hypothesis can be more directly tested.

We compute $T_c$ in the presence of impurities
from the standard mean field formalism \cite{AM}
in which the electron self-energy is
solved self-consistently from the single-exchange graph.
This approach is justified if Migdal's theorem \cite{Migdal}
applies, which is the case in conventional, phonon-mediated
superconductors \cite{Scalapino}, but it is not clear whether
a similar results holds in
the high-temperature superconductors \cite{vertex}.
Nonetheless, we follow other authors \cite{MBP,Lenck,Hotta,MP,Ueda,RULN}
and assume that this result holds in what follows.

In most of our calculations, we employ
the standard set of approximations to this mean-field theory in order
to simplify the resulting Eliashberg equations \cite{AM,Scalapino}.
These simplified equations are discussed in more detail in
Ref. \cite{RULN}.
The central approximation in this approach is
to limit the wavevectors of the electron self-energy and pairing
potential to the Fermi surface, so
we refer to the solution of these approximate equations as the
Fermi-surface-restricted solution.
For comparison, we have also solved the Eliashberg equations
without these approximations on two-dimensional
lattices of small size (64 x 64, 32 x 32)
using recently developed fast Fourier transform techniques
\cite{MBP,MP,Serene}.
We refer to the solution of these equations as the exact solution.

Of the models of d-wave superconductors available
\cite{MBP,Lenck,MP,Ueda,RULN,Bulut},
we employ a spin-fluctuation-mediated pairing
interaction with a spectrum constrained by neutron scattering in
$\rm YBa_2Cu_3O_{6.7}$ \cite{RULN}.
Solving the Fermi-surface-restricted equations
with the pairing potential and band structure discussed in
Ref. \cite{RULN} gives us the critical temperature as a
function of the bare impurity scattering rate $\tau_{imp}^{-1}$.
(Throughout this paper, we set $\hbar = k_B = 1$).
We compute $T_c$ vs. $\tau_{imp}^{-1}$ for a variety of critical
temperatures in the absence of impurities $T_{c0}$ and in
both the weak- (inelastic scattering ignored) and
strong- (inelastic scattering included) coupling cases.
Although we vary the $T_{c0}$'s in this paper, constraints on the
electron-spin fluctuation coupling constant imposed by ac
conductivity require that $T_{c0}$ = 7.2 K in this model \cite{RULN}.
For comparison, we also compute the impurity-induced $T_c$ suppression
from the exact Eliashberg equations for the Monthoux-Pines
spin-fluctuation model \cite{MBP,MP} with a coupling constant chosen
so that $T_{c0}$ = 100 K.

We will show that these numerical results can be approximately
collapsed to a universal curve.
One can deduce that the form of this curve for d-wave superconductors is
\begin{equation}
- \ln \left( \frac{T_c}{T_{c0}} \right)
     = \psi \left( \frac{1}{2} + \frac{\alpha T_{c0}}{2 \pi T_c} \right)
       - \psi \left( \frac{1}{2} \right) , \label{eq:AG}
\end{equation}
where $\psi(z)$ is the digamma function,
$\alpha$ = \mbox{$1 / (2 (1 + \lambda_Z) \tau_{imp} T_{c0})$}
is the strong-coupling pair-breaking parameter, and
$ \lambda_Z $
  = 1 - \mbox{$\langle
             Im [\Sigma^{boson} ({\bf k}, i\omega_0) ] \rangle /
             \omega_0 $}.
In this last equation, $\Sigma^{boson} ({\bf k}, i\omega_0)$ refers
to the self-energy of the pairing boson at the lowest fermionic
Matsubara frequency $\omega_0 = \pi T_{c0} $, and the angle brackets
denote an average over the Fermi surface.
We note that, for an s-wave
superconductor with magnetic impurities,
Eq. (\ref{eq:AG}) would still apply, but without the
factor of two in $\alpha$.
We note that individual aspects of this formulation have
appeared in the works of other authors \cite{MSV,AG,AM};
in particular, this form of the pair-breaking parameter can be
found in Ref. \cite{MSV}.

In order to make contact with experiment, we
represent the pair-breaking parameter $\alpha$ in
terms of the planar residual resistivity $\rho_0$.
{}From the Kubo formula under the standard assumptions that
the pairing boson energy
is small compared to the electronic energies and that vertex corrections
are not important \cite{Mahan}, the real part of the low frequency
residual electrical conductivity is given by
\begin{equation}
 {\rm Re} \, \sigma (\omega)
   =  \frac{\omega_{pl}^{*2}}{4 \pi} \,
     \frac{\tau_{imp}^*}{1 + \omega^2 \tau_{imp}^{*2}}. \label{eq:Drude}
\end{equation}
In this equation, $\omega_{pl}^{*2} = \omega_{pl}^{2} / (1 + \lambda)$
is the renormalized plasma frequency which is measured experimentally,
\begin{equation}
  \frac{1}{\tau_{imp}^{*}}
     = \frac{1}{1 + \lambda} \, \frac{1}{\tau_{imp}}
\end{equation}
is the renormalized scattering rate due to impurities, and
$\lambda = - \left. \langle \partial {\rm Re}
  \Sigma^{boson} ( {\bf k}, \omega) /
  {\partial \omega} \rangle \right|_{\omega = 0} $
is the mass renormalization parameter due to the pairing bosons.
At zero temperature, $\lambda$ is equal to the
inelastic scattering parameter from the Eliashberg equations
$\lambda_Z$.
Since the characteristic energy of the pairing boson is
low compared to other electronic energy scales (i.e., of the order
of $T_c$), we find that $\lambda \cong \lambda_Z$ to within 5-10 \%
at $T_c$.
Combining this result with Eq. (\ref{eq:Drude}) and the definition
of the pair-breaking parameter $\alpha$, we see that
\begin{equation}
 \rho_0
    \cong \frac{4 \pi}{\omega_{pl}^{*2}} \,
          2 T_{c0} \, \alpha. \label{eq:rho0}
\end{equation}
We emphasize that Eq. (\ref{eq:rho0}) allows us to
predict the $T_c$ suppression as a function
of planar residual resistivity from the experimentally observed plasma
frequency $\omega_{pl}^{*2}$ and the critical temperature $T_{c0}$
{\it independently} of any microscopic model.

Having set up the formalism for our calculations, we will now discuss
the results.
Fig. 1 shows $T_c / T_{c0}$ vs. scattering rate
from non-magnetic impurities for a selection of
$T_{c0}$'s in the model d-wave superconductor of Ref. \cite{RULN}.
Since $T_{c0}$ is inversely proportional to the coherence length $\xi$,
the different curves in each figure correspond to different
$\xi$'s.
Both weak- (Fig. 1(b)) and
strong- (Fig. 1(a)) coupling calculations show that, as $T_{c0}$
increases ($\xi$ decreases), the superconductor becomes less
sensitive to the presence of non-magnetic impurities.
In addition, by comparing curves in
Figs. 1(a) and 1(b) with the same $T_{c0}$, we see that the
strong-coupling
curves (which include the effects of inelastic scattering) are less
sensitive to impurities than the weak-coupling
results \cite{BigTc}.  These trends conform to the qualitative expectations
discussed in the introduction.

To confirm the validity of our analytical results for the
scaling function Eq. (\ref{eq:AG}), we plot in Fig. 2 the
numerical data in Fig. 1 in terms of the scaled variables
$T_c / T_{c0}$ and $\alpha$.
As can be seen from the figure, all of the scaled curves
cluster near the scaling function.
Note that the data in Fig. 2 include
both weak- and strong-coupling results;
moreover, even the curve from Fig. 1(a)
with nearly the maximum achievable $T_{c0}$ = 50 K falls near
the scaling curve. In the inset to Fig. 2, we display the
impurity-induced suppression of a 100 K d-wave superconductor
in the Monthoux-Pines model computed from the exact Eliashberg
equations.
When plotted in terms of $T_c / T_{c0}$ and $\alpha$, this model
also produces a $T_c$ suppression which is close to the scaling
function Eq. (\ref{eq:AG}).

Having established the validity of the scaling law, we use this
relation to predict the response of a 90 K superconductor to
non-magnetic impurities.
In Fig. 3, the shaded region corresponds to the
d-wave $T_c$ computed from Eqs. (\ref{eq:AG}) and (\ref{eq:rho0})
for plasma frequencies $\omega_{pl}^{*}$ ranging from 1.1
to 1.4 eV as a function of the planar residual resistivity
$\rho_0$.
These plasma frequencies are chosen to reflect the range of
experimental uncertainty in $\omega_{pl}^{*}$ in
$\rm YBa_2Cu_3O_7$ \cite{Timusk}.
For comparison, Fig. 3 also shows the expected response
of an s-wave superconductor
to non-magnetic impurites (dashed line);
in accordance with Anderson's theorem
\cite{Anderson}, these impurities have no effect on $T_c$.

In addition to the uncertainty in the plasma frequency, the
accuracy of the prediction in Fig. 3 is affected by the applicability of the
scaling law and the assumption of structureless impurities.
{}From Fig. 2, one can see a systematic trend away from the scaling
curve as $T_{c0}$ increases.  For a 90 K superconductor, this
error amounts to roughly a 20 \% correction to the horizontal scale
in Fig. 3, which is about the
same magnitude as the uncertainty in the plasma frequency.
The effect of ignoring the wavevector dependence
of the impurity scattering matrix element is more difficult to estimate.
In extreme cases, impurities with wavevector structure could alter
the scale on the horizontal axis by a factor of two \cite{MSV}.
Even considering these caveats,
if the superconducting order parameter has d-wave symmetry,
then the prediction in Fig. 3  should give the
correct scale for the residual resistivity at which significant
depression of the critical temperature will occur in
$\rm YBa_2Cu_3O_7$ .

Experimentally, one can estimate the residual resistivity $\rho_0$ by
extrapolating the measured linear resistivity vs. temperature
for temperatures T greater than $T_c$ to T = 0.
It is found that this extrapolated planar $\rho_0$ is about
20 $\mu\Omega$-cm. in high-quality twinned crystals of $\rm YBa_2Cu_3O_7$
and is roughly - 20 $\mu\Omega$-cm. in the best untwinned crystals.
A negative extrapolated value of $\rho_0$ means that the
resistivity in the absence of superconductivity could
not remain linear all the way down to zero
temperature, but must turn over to some higher power law.
The $T_c$'s of twinned and untwinned crystals are about the same,
despite the fact that the change in residual resistivity is of order
20 $\mu\Omega$-cm. between the two types of samples.
By contrast, if $\rm YBa_2Cu_3O_7$ were a d-wave
superconductor, then, according to Fig. 3, $T_c$ would be strongly
suppressed.

This naive argument against the d-wave scenario is not strictly
correct, since
our calculations have focused entirely
on a single $\rm CuO_2$ plane, which implies that the
residual resistivity in Fig. 3 represents only
the planar contribution to $\rho_0$ and does not include
the contribution from the $\rm CuO$ chains.
For this reason,
one can not simply take existing measurements of
$T_c$ and $\rho_0$ in polycrystalline or untwinned samples and
infer unambiguous evidence for
or against d-wave superconductivity; more systematic experiments
are required.

Currently, two methods could be used to perform such systematic
experiments:  substitution and irradiation.
Substitutional studies \cite{Markert,Chien,Xiao}
are able to introduce defects mainly in the $\rm CuO_2$ planes, as
required, but find that the induced defects are generally magnetic.
Abrikosov-Gor'kov-like behavior given by Eq. (\ref{eq:AG})
is then to be expected and
provides no qualitative distinction between s- and d-wave pairing.
Alternatively, defects can be induced by irradiation
\cite{White,Vichery,Hofmann}, but the location of these defects
is often uncontrolled.
Recent experiments \cite{Giap}, though, suggest that
low-energy electrons may preferentially disorder the Cu or O in
the planes, which would enable a comparison of irradiation data with
our predictions.
Current irradiation studies show that $T_c$ decreases linearly with
fluence, as one would expect for weak disorder in the scaling curve of
Eq.~(\ref{eq:AG}), but it is not known how the fluence is
correlated to the residual resistivity or whether the induced
defects carry magnetic moments or not.
Thus, although it is not yet possible to make a quantitative
comparison of our results with current experiments,
such a comparison should be
possible with future systematic measurements.
We hope that the present paper will serve as a stimulus
for such experiments.

\acknowledgements

This work was supported by NSF-STC-9120000.  HBS was also
supported by
NSF-DMR-8913878 and NSF-DMR-9215123
and MRN by the U.~S.~Department of Energy,
BES-Materials
Sciences, under Contract \#W-31-109-ENG-38.
We also acknowledge helpful conversations with G. W. Crabtree,
J. Giapintzakis, D. M. Ginsberg, W.-K. Kwok, S. L. Cooper,
and T. Timusk.

\figure{Critical temperature $T_c$ normalized by the critical
temperature in the absence of impurities $T_{c0}$ vs. the bare
impurity scattering rate
$\tau_{imp}^{-1}$ in meV for the model of Ref. \cite{RULN} in the (a)
strong-coupling and (b) weak-coupling cases calculated in the
Fermi-surface-restricted Eliashberg formalism.  The $T_{c0}$'s are
7.2 K
(circle), 31.6 K (box), and 50.2 K (triangle); solid symbols denote
strong-coupling results and empty symbols denote weak-coupling
results.
The solid lines are to guide the eye.\label{tcvsrho}}
\figure{Normalized critical temperature $T_c / T_{c0}$ vs.
pair-breaking parameter
$\alpha$ = \mbox{$1 / 2 (1 + \lambda_Z) \tau_{imp} T_{c0}$} for
the data in
Fig. 1.  Plot symbols are the same as in Fig. 1 with the solid symbols
denoting
strong-coupling calculations and the open symbols denoting weak-
coupling
calculations.  For comparision, the analytic form of the scaling
function [Eq. (\ref{eq:AG})] is plotted as a plain solid line.
Inset:  Normalized critical temperature vs. pair-breaking parameter
for the model of Refs. \cite{MBP} and \cite{MP}
calculated from the exact
Eliashberg equations with the coupling chosen so that
$T_{c0}$ = 99.4 K. \label{scaling}}
\figure{Prediction of $T_c$ as a function of the in-plane residual
resistivity $\rho_0$ due to non-magnetic impurities
in a 90 K superconductor with
a d-wave (shaded area) or an s-wave
(dashed line) order parameter.  The d-wave curve is computed from
the
generalized Abrikosov-Gor'kov form [Eqs.~(\ref{eq:AG}) and
(\ref{eq:rho0})] for experimental plasma frequencies $\omega_{pl}^{*}$
between 1.1 and 1.4 eV, and the
s-wave curve is
simply a straight line due to Anderson's theorem.\label{nonmag}}


\begin{references}
\bibitem{BigTc}Although the $T_{c0}$ = 50 K curve in Fig. 1(a) appears to be
very insensitive to impurity content, this appearance is an artifact
of plotting $T_c / T_{c0}$ against the bare lifetime; when
plotted against the (observable) renormalized lifetime, the
decrease in $T_c$ for all the curves is similar.
This behavior is demonstrated explicitly in the scaling curve
(Fig. 2).
\end{references}
\end{document}